# GrifFinNet: A Graph-Relation Integrated Transformer for Financial Predictions

Chenlanhui Dai[1,2], Wenyan Wang[1,3], Yusi Fan[1,3], Yueying Wang[1,3], Lan Huang[1,3], Kewei Li[1,3,#], Fengfeng Zhou[1,3,#].

1 College of Computer Science and Technology, Jilin University, Changchun, China, 130012.

2 School of Economics, Jilin University, Changchun, China, 130012.

3 Key Laboratory of Symbolic Computation and Knowledge Engineering of Ministry of Education, Jilin University, Changchun, China, 130012.

# Correspondence may be addressed to Fengfeng Zhou: FengfengZhou@gmail.com or ffzhou@jlu.edu.cn . Lab web site: http://www.healthinformaticslab.org/ . Phone: +86-431-8516-6024. Fax: +86-431-8516-6024. Correspondence may also be addressed to: Kewei Li, kwbb1997@gmail.com.

# Abstract

Predicting stock returns remains a central challenge in quantitative finance, transitioning from traditional statistical methods to contemporary deep learning techniques. However, many current models struggle with effectively capturing spatio-temporal dynamics and integrating multiple relational data sources. This study proposes GrifFinNet, a Graph-Relation Integrated Transformer for Financial Predictions, which combines multi-relational graph modeling with Transformer-based temporal encoding. GrifFinNet constructs inter-stock relation graphs based on industry sectors and institutional ownership, and incorporates an adaptive gating mechanism to dynamically integrate relational data in response to changing market conditions. This approach enables the model to jointly capture spatial dependencies and temporal patterns, offering a comprehensive representation of market dynamics. Extensive experiments on two Chinese A-share indices show that GrifFinNet consistently outperforms several baseline models and provides valuable, interpretable insights into financial market behavior. The code and data are available at: https://www.healthinformaticslab.org/supp/.

# Introduction

Stock return prediction has been a crucial task in quantitative finance. Unlike static analysis, the stock market is inherently complex and subject to sudden shifts, and the development of dynamic models is essential to adapt to the evolving market conditions [1]. The market is influenced by numerous factors, such as economic developments, policy changes, and news events, while its fluctuations are also shaped by historical data [2].

The study of stock returns prediction traces back to traditional statistical models. The Capital Asset Pricing Model (CAPM) [3] provided a foundational understanding of systematic risk and expected returns. This approach was extended by Fama &

French's three-factor [4] and five-factor models [5], which introduced profitability and investment factors alongside original market, size, and value factors to explain cross-sectional variations in stock returns. Linear models, such as GARCH [6], were also developed to model time-varying volatility, while retaining linear assumptions based on the efficient market hypothesis. While these models provided a solid baseline, they have gradually shown limitations in addressing the complexity of modern financial markets, especially due to their reliance on linear assumptions and the efficient market hypothesis. This restricts their ability to capture non-linear relationships and the dynamic behaviors inherent in financial systems.

With the rapid development of computational technology, machine learning techniques have gained prominence in stock return prediction. Machine learning algorithms such as Support Vector Machines (SVM) [7], Random Forests [8], and XGBoost [9] are capable of handling high-dimensional features and non-linear relationships, with improved prediction accuracy over traditional statistical models [10]. However, these methods have notable shortcomings, particularly in modeling temporal dependencies and fully leveraging the time-series characteristics of financial data.

Deep learning models, including Recurrent Neural Networks (RNN) [11], Long Short-Term Memory (LSTM) networks [12], and Transformers [13], have advanced the effective modeling of temporal dependencies and the capture of long-term trends in stock prices. At the same time, Graph Neural Networks (GNNs) have gained interests across various domains, showing particular promise in financial applications. The Graph Convolutional Network (GCN) [14] and Graph Attention Network (GAT) [15] have established the foundation for processing graph-structured data. More recently, methods combining heterogeneous graph neural networks with attention mechanisms, such as those proposed by [16], have demonstrated substantial power in handling complex relational data. In stock modeling, these methods offer a compelling approach to capturing inter-stock relationships, providing valuable insights for modeling the complex structural dependencies in financial markets. By leveraging graph-based structures, these models can dynamically capture the interconnections between stocks, enhancing stock return prediction and improving the ability to model intricate dependencies within the market.

This study presents GrifFinNet, a graph–relation integrated Transformer for stock return prediction that unifies multi-relational spatial modeling with temporal encoding. GrifFinNet (i) constructs two explicit inter-stock graphs (industry affiliation and institutional co-ownership) to encode economically grounded dependencies; (ii) introduces an adaptive gating fusion module that dynamically weights relation types in response to market regimes; and (iii) applies a Transformer encoder over the fused representations to capture long-range temporal structure. The problem is formulated as cross-sectional return ranking with robust preprocessing (winsorization and Z-score normalization), and the model is trained end-to-end. Extensive experiments on CSI300 and CSI800 benchmark datasets show consistent gains over sequential, tree-based, and graph baselines across ranking and portfolio metrics. The ablation experiments confirm the necessity of multi-relation modeling, gating, and temporal encoding. Analysis of learned gate dynamics further provides interpretable evidence of regime sensitivity, and underscores the practical relevance of the proposed design.

# Related Work

A central challenge in stock market modeling is to jointly capture multiple inter-stock relations together with temporal dynamics to improve predictive performance [17].

## Spatio-Temporal Modeling for Stock Prediction

Stock prices reflect both their own histories and cross-sectional influences from related securities; hybrid integration methods aim to bridge these sources of information. Early spatio-temporal graph neural networks incorporate predefined ties—e.g., shareholder/corporate relations [18] and industry sectors [19], [20]—to encode economically meaningful dependencies, but their adaptability in highly volatile regimes is limited. To address this, subsequent work learns dynamic relations, including latent conceptual links [21] and hypergraph-based higher-order structures [22], improving robustness to shifting market conditions while retaining temporal sensitivity.

## Multi-Relation and Gating Mechanisms

A parallel line of research explores the integration of temporal modeling with advanced gating architectures. The MASTER framework [23] furthered this idea by introducing a market-guided gating mechanism that adaptively re-weights features according to prevailing market states, and the framework enhanced robustness across different regimes. A similar study [24] proposed a hybrid approach combining Multivariate Empirical Mode Decomposition (MEMD) with an Aquila optimizer–enhanced LSTM. In this design, a gating component regulates the flow of decomposed signal components, and enables effective multi-source fusion and yielding high predictive accuracy.

In the area of multi-relation modeling, the HATR-I model [25] introduces hierarchical adaptive temporal-relational interactions，which grasps short- and long-term transition regularities of stock dynamics based on cascaded dilated convolutions and gating paths.[26] introduced the LSR-IGRU framework, which extends GAT with an improved GRU [27] and incorporates a long-short-term relation modeling mechanism. By jointly constructing industry-based and overnight price-change relation graphs, LSR-IGRU effectively captures both structural and temporal dependencies among stocks, marking an important step toward comprehensive multi-level relational modeling.

## Limitations and Our Approach

Despite advances in spatio-temporal integration and multi-relation construction, two major limitations remain. First, most existing methods treat spatial and temporal modules as loosely connected components. Temporal models therefore capture sequential patterns without accounting for evolving inter-stock relations, while spatial models neglect the concurrent temporal dynamics. This separation leads to suboptimal joint representation learning. Second, current gating mechanisms primarily operate at the feature-selection level. They lack the capacity to adaptively fuse heterogeneous relation types or dynamically adjust their relative importance in response to shifting market conditions.

To address these gaps, we propose a graph-relation integrated Transformer framework GrifFinNet for financial prediction, which is inspired by recent work on

adaptive fusion in relational modeling [28]. Our main contributions are as follows:

- **Comprehensive multi-relation construction**: We build a unified graph that jointly incorporates industry-based and institutional-based relations, and captures diverse inter-stock dependencies.
- **Dual gating fusion mechanism**: We design a gating framework that dynamically weights and fuses heterogeneous relations. By adaptively re-allocating attention between industry and institutional graphs according to current market conditions, the model emphasizes the most informative connections in real time.
- **Tight spatial-temporal integration**: Unlike prior approaches that process spatial and temporal features separately, GrifFinNet fuses relational representations through the dual gating mechanism and directly encodes them with a Transformer. This end-to-end design enables the simultaneous optimization of graph-based spatial dependencies and temporal price dynamics within a single architecture.

## Materials and Methods

### Problem Definition

Building on prior work in stock market analysis [20], [22], [29], [30], this study focuses on predicting stock returns rather than absolute price levels. Returns directly reflect investment profitability and are therefore more relevant for portfolio allocation and risk management decisions.

For each stock $i \in \{1, 2, \ldots, N\}$ in the universe, we collect a 221-dimensional feature vector $F_i^t \in \mathbb{R}^{221}$ at each time step as input. The prediction target is the corresponding stock return. To ensure robustness, the labels $y_i^t$ undergo winsorization to remove extreme outliers, followed by Z-score normalization. This normalization converts absolute returns into relative ranking information, which is consistent with practical investment objectives since investors are generally more concerned with the relative performance of stocks rather than their raw return values.

Formally, the input is represented as a temporal feature matrix, while the output is the normalized return prediction vector $\hat{y}_i^t$, where $i \in \{1,2,\ldots,N\}, t \in \{1,2,\ldots,T\}$, $N$ denotes the number of stocks and $T$ the length of the time series. This formulation aligns the task with real-world investment strategies and provides a standardized framework for evaluating model effectiveness.

## Overall Architecture

The overall framework of **GrifFinNet** is shown in Figure 1. The model is composed of four tightly integrated modules:

- **Graph Relational Attention Mechanism** constructs industry- and institution-based graphs to explicitly model economically grounded inter-stock relations, and ensures the preservation of structural dependencies.
- **Adaptive Gating Fusion Mechanism** dynamically balances the contributions of heterogeneous relations, and allows the model to emphasize the most relevant information under different market regimes.
- **Temporal Transformer Encoder** processes the fused relational representations to capture long-range temporal dependencies, which are critical in financial time series with delayed effects.
- **Prediction Module** aggregates temporal features through global pooling and produces return forecasts with a multilayer perceptron head. This strategy enables end-to-end learning from raw features to actionable outputs.

This design allows GrifFinNet to unify spatial and temporal modeling within a single architecture, and addresses the limitations of prior methods that process them separately.

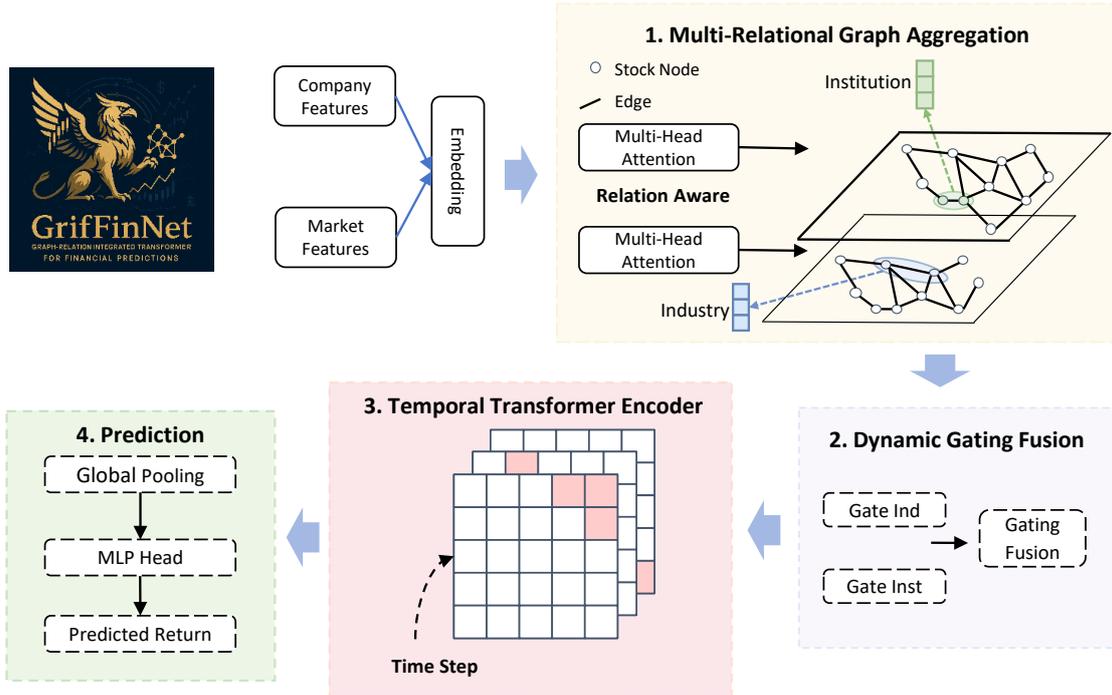

**Figure 1. Architecture of GrifFinNet.** The framework integrates graph relational attention, adaptive gating fusion, temporal encoding, and prediction modules for stock return forecasting.

## Feature Processing and Representation Learning

In stock return prediction, feature sets are inherently heterogeneous. Following the dataset design of [23], we categorize features into company-specific indicators (158 dimensions) and market-level indicators (63 dimensions). To accommodate these differences, the two categories are processed separately through linear transformations:

$$\begin{aligned} X_{company} &= \text{Linear}(F_{company}) \in \mathbb{R}^{N \times T \times d_{company}} \\ X_{market} &= \text{Linear}(F_{market}) \in \mathbb{R}^{N \times T \times d_{market}} \end{aligned} \quad (1)$$

where $F_{company} \in \mathbb{R}^{N \times T \times 158}$ and $F_{market} \in \mathbb{R}^{N \times T \times 63}$ denote the raw feature matrices. $T$ represents the temporal sequence length, $d_{company} = 158$ and $d_{market} = 63$ represent the output dimensions.

To provide the model with a notion of temporal order and positional context, we apply sinusoidal positional encoding to the processed features for the compatibility

with subsequent relational modeling:

$$\begin{aligned} \text{PE}_{(pos,2i)} &= sin(pos/10000^{\frac{2i}{d_{model}}}) \\ \text{PE}_{(pos,2i+1)} &= cos(pos/10000^{\frac{2i}{d_{model}}}) \end{aligned} \quad (2)$$

where $pos$ is the time-step index, $i$ is the dimension index, and $d_{model}$ is the dimension of the model embedding features. This encoding scheme preserves sequential dependencies and enables effective temporal alignment across heterogeneous features.

The processed company and market features are then concatenated and combined with positional encodings to form the input representation $X$ for subsequent relational modeling:

$$X = \text{Concat}(X_{\text{company}}, X_{\text{market}}) + PE \in \mathbb{R}^{N \times T \times d_{\text{model}}} \quad (3)$$

## Graph Relational Attention Mechanism

In financial markets, stocks are influenced by complex interdependencies that cannot be fully captured by single-modality models. To address this, we construct two types of relation graphs, i.e., **industry-based** and **institution-based**, to explicitly encode structured relationships among stocks.

Using the Tushare financial data interface [31], we obtain industry and institutional affiliation information for CSI300 and CSI800 constituents. The industry relation matrix $A_{ind}$, where $A_{ij,ind}$ encodes the association strength between stocks $i$ and $j$, reflecting that firms within the same sector are typically exposed to similar macroeconomic factors, regulatory changes, and sectoral trends. In parallel, the institutional relation matrix $A_{inst}$, where $A_{ij,inst}$ quantifies co-ownership effects, capturing the coordinated influence of institutional investors on stock price movements between stocks $i$ and $j$. These yield the industry relationship graph $G_{ind}$ and the institutional relationship graph $G_{inst}$:

$$G_{ind} = (S, E_{ind}, A_{ind}) \,;\, G_{inst} = (S, E_{inst}, A_{inst}) \quad (4)$$

where $S$ is the set of stock nodes, and $E_{ind}$ and $E_{inst}$ are the respective edge sets representing association patterns.

We integrate these relations into the attention computation. Given the feature matrix $X \in \mathbb{R}^{N \times d}$, the (query, key, and value) projections are defined as $Q = XW_Q$, $K = XW_K$, and $V = XW_V$. We denote the relation matrix as $R \in \{A_{ind}, A_{inst}\}$ and the attention is calculated as:

$$\text{Attention}(Q, K, V, R) = \text{softmax}\left(\frac{QK^T}{\sqrt{d_k}} + \alpha R\right) V \tag{5}$$

where $\alpha$ is a learnable scaling factor, and $d_k$ is the dimension of the key vector. This ensures that attention weights between stock pairs are strengthened when domain-specific relationships exist. Finally, for each relation type $r \in \{ind, inst\}$, we compute the hidden state representation:

$$X_r = \text{Attention}(Q, K, V, R) \tag{6}$$

This relation-enhanced attention mechanism effectively embeds structured industry and institutional information into the attention process, and enables the model to capture multi-level interactions among stocks during feature aggregation.

## Adaptive Gating Fusion Mechanism

In financial markets, the relevance of different types of stock relations varies with market conditions. To capture this variability, we introduce an adaptive gating fusion mechanism that dynamically regulates the contributions of industry- and institution-based relations. This design enables the model to prioritize the most informative connections under changing environments.

For each stock $i$ and relation type $r \in \{ind, inst\}$, the gate values are computed as:

$$g_{i,r} = \sigma(W_{g,r} X_i + b_{g,r}) \tag{7}$$

where $\sigma(\cdot)$ is the sigmoid function, and $W_{g,r}$ and $b_{g,r}$ are learnable parameters.

The gate values lie in $[0,1]$, thereby allowing continuous control over the information flow contributed by each relation type.

The fused representation is then obtained by:

$$X_{fused} = \frac{g_{i,ind} \odot X_{i,ind} + g_{i,inst} \odot X_{i,inst}}{2}$$
$$X_{out} = W_{proj}X_{fused} + X \qquad (8)$$

where $\odot$ denotes element-wise multiplication (Hadamard product), and $W_{proj}$ ensures dimensional alignment with the residual connection. This residual structure preserves feature integrity while enabling the gates to selectively amplify or suppress relational signals according to market state.

## Temporal Transformer Encoding

Stock prices display strong temporal dependencies, and a good predictive model needs to capture both short- and long-range correlations. To this end, we employ a Transformer encoder to process the fused relational features.

The fused relational features $X_{out}$ from equation (8) serve as the initial input $X^{(0)}$. At each layer $l$, the encoder performs multi-head self-attention followed by a position-wise feed-forward network, both wrapped with residual connections and normalization:

$$X^{(l)} = X^{(l-1)} + \text{MultiHeadAttention}\left(\text{LayerNorm}(X^{(l-1)})\right)$$
$$X^{(l)} = X^{(l)} + \text{FFN}(\text{LayerNorm}(X^{(l)})) \qquad (9)$$

where FFN() is the feed-forward network, MultiHeadAttention() represents the multi-head attention mechanisms and $l$ denotes the layer index.

The nonlinear feed-forward transformation increases the representational power of the encoder, and enables it to capture complex feature interactions. Meanwhile, residual connections and layer normalization stabilize training, mitigate gradient degradation, and accelerate convergence. By stacking these layers, the Transformer encoder effectively captures long-term temporal dependencies while maintaining stable gradient flow. These characteristics make the architecture particularly well suited to financial time series, where lagged effects and extended temporal correlations are prevalent.

## Prediction and Training

After temporal and relational feature extraction, we obtain sequences containing rich contextual information. To form a unified representation, we apply average pooling across all time steps: $X_{global} = (\sum_{t=1}^{T} X_t^{(L)})/T$, where $T$ is the length of the time series and $X_t^{(L)}$ denotes the output of the $L^{th}$ Transformer layer at time step $t$. The pooled representation is passed through a multi-layer prediction head and generates the final return predictions:

$$\hat{y} = w_{out} \text{ReLU}(W_{hidden} X_{global} + b_{hidden}) + b_{out} \tag{10}$$

We train the model with a mean squared error (MSE) loss augmented with L2 regularization:

$$Loss = \frac{1}{N} \sum_{i=1}^{N} (y_i - \hat{y}_i)^2 + \lambda |\theta|_2^2 \tag{11}$$

where $\lambda$ controls the strength of regularization and $\theta$ represents the trainable parameters. The L2 term mitigates overfitting and improves generalization.

For optimization, we adopt AdamW, which integrates decoupled weight decay for improved stability, together with the OneCycleLR learning rate scheduler to accelerate convergence and reduce sensitivity to initialization. These design choices ensure efficient and robust training across large-scale financial datasets.

## Experiments

### Datasets

We evaluate GrifFinNet on two widely used Chinese equity benchmarks CSI300 and CSI800. Following the dataset protocol of MASTER [23], the CSI300 dataset contains the top 300 stocks and the CSI800 dataset the top 800 stocks, both drawn from the Shanghai and Shenzhen stock exchanges. The sample period spans January 2008 to December 2022, which is partitioned into training, validation, and test sets with cut-off dates aligned to real-world trading timelines (Table 1) [23].

Feature construction is based on the Alpha158 indicators [32], a standard factor set for quantitative investment research. To represent market-level dynamics, we additionally construct 63 market features derived from CSI300 and CSI800 index data. All features undergo standardized preprocessing following the MASTER and Qlib pipelines to ensure comparability across models.

**Table 1. Overview of CSI300 and CSI800 datasets used in experiments.** Each dataset spans 2008-2022 and is divided into training, validation, and testing subsets. Features are constructed from Alpha158 indicators and enriched with market-level variables.

| Dataset | Subset | Time Range | Sample Size |
| --- | --- | --- | --- |
| **CSI300** | Training | 2008-01-01-2020-03-31 | 856,247 |
|  | Validation | 2020-04-01-2020-06-30 | 20,040 |
|  | Testing | 2020-07-01-2022-12-31 | 185,701 |
| **CSI800** | Training | 2008-01-01-2020-03-31 | 1,757,733 |
|  | Validation | 2020-04-01-2020-06-30 | 53,419 |
|  | Testing | 2020-07-01-2022-12-31 | 495,122 |

## Baselines

To assess the effectiveness of GrifFinNet, we compare it against a diverse set of baseline models spanning sequential architectures, graph-based methods, and traditional machine learning approaches, as well as several state-of-the-art (SOTA) models.

We include three widely used sequential models: LSTM [12] captures long-term dependencies in time series; GRU [27] is a streamlined variant of LSTM designed to balance predictive power with computational efficiency; and Transformer [13] models complex temporal correlations through self-attention mechanisms.

To capture inter-stock dependencies, we evaluate graph-based models like GAT [15],

which encodes stock representations and aggregates neighborhood information using attention weights; GCN [14], which propagates information via spectral graph convolutions; and DTML[33], which leverages attention mechanisms to model dynamic temporal relationships across stocks.

Two traditional machine learning algorithms are compared with the proposed GrifFinNet framework. XGBoost [9] is a gradient-boosted tree ensemble that remains a strong benchmark for tabular financial data. Random Forest [8] is a classical ensemble method based on bagging and decision trees.

Three recently-published SOTA models are also evaluated with GrifFinNet. MASTER [23] is a market-guided Transformer that alternates intra-stock and inter-stock attention to capture short-term and cross-time dependencies, with gating informed by market conditions. A lightweight MLP-Mixer framework StockMixer [34] models indicator-, temporal-, and stock-level correlations through multi-scale patches and information flow. A probabilistic dynamic-factor model based on variational autoencoders FactorVAE [35] was proposed to extract robust latent factors from noisy data while jointly predicting returns and risks.

## Performance Metrics

To evaluate predictive and investment performance, we adopt both ranking-based metrics and portfolio-based metrics, complemented by additional measures for risk–return analysis.

We employ four commonly used ranking measures in quantitative finance: Information Coefficient (IC) is calculated by the Pearson correlation between predicted and realized returns for measuring linear alignment. Rank Information Coefficient (RankIC) measures the Spearman correlation to capture rank-order consistency. ICIR and RankICIR measure the information ratios of IC and RankIC to their respective standard deviations, and are computed over rolling windows to reflect the stability of predictive signals.

Portfolio metrics are calculated to assess practical investment outcomes, and we implement a daily trading simulation that selects the top 30 stocks ranked by predicted returns. From this strategy we compute: Excess Annualized Return (AR) by the expected annualized return relative to a benchmark for the quantified profitability, and Information Ratio (IR) by the ratio of excess return to its volatility for the quantification of risk-adjusted efficiency.

For component-level evaluation, we conduct risk–return analysis using two additional metrics. The Sharpe Ratio quantifies excess return per unit of risk $SR = E[R - R_f]/\sigma[R - R_f]$, where $R_f$ is the risk-free rate [30]. This standardized ratio facilitates cross-model comparison of risk-adjusted performance. The Cumulative Return is defined as the total return trajectory over the test horizon, providing a direct visualization of long-term investment outcomes.

Together, these measures complement ranking and portfolio metrics, and offer a balanced framework that integrates statistical accuracy, practical profitability, and risk management.

## Implementation Details

We implement GrifFinNet using the PyTorch framework (version 1.13.1) under Python 3.11 and re-implement all baseline models to ensure consistency across experiments.

Sequential models (LSTM, GRU, and Transformer) are implemented with PyTorch. For graph-based approaches, we develop custom implementations of GAT, GCN, and DTML to encode inter-stock relationships. Traditional machine learning baselines (XGBoost and Random Forest) are constructed using the XGBoost library (version 1.7.3) and scikit-learn (version 1.2.2), respectively. Recently-published SOTA methods, including MASTER, FactorVAE, and StockMixer, are re-implemented according to their original papers to guarantee faithful reproduction.

Following MASTER [23], we tailor hyperparameters to dataset characteristics and determine optimal values through parameter tuning. The learning rate is fixed at $10^{-4}$, selected from candidates $\{10^{-3}, 10^{-4}, 10^{-5}\}$. Model dimensions are set to 256 for CSI300 and 512 for CSI800, reflecting differences in dataset scale. Consistent with MASTER, the coefficient $\beta$ is set to 5 for CSI300 and 2 for CSI800. Sensitivity studies further identify the best configuration as: one Transformer layer, eight attention heads, and a dropout rate of 0.15. Detailed results of the parameter sensitivity analysis are provided in the section "Parameter Sensitivity Analysis".

All experiments are conducted on a server equipped with an Intel® Xeon® E5-2686 v4 CPU (18 cores, 36 threads), 64 GB memory, and an NVIDIA RTX 3090 GPU (24 GB VRAM). To ensure robustness, each experiment is repeated five times with different random seeds, and average results are reported.

# Results and Discussion

## Model Results and Comparison

We evaluate GrifFinNet on both the CSI300 and CSI800 datasets, with results summarized in Table 2.

Table 2. Performance of GrifFinNet on CSI300 and CSI800 datasets.

| Dataset | IC | ICIR | RIC | RICIR | AR | IR |
|---|---|---|---|---|---|---|
| CSI300 | 0.081 | 1.844 | 0.081 | 1.822 | 1.278 | 3.937 |
| CSI800 | 0.097 | 1.837 | 0.097 | 1.817 | 0.913 | 1.198 |

Table 3. Comparison with baseline models on CSI300.

| Model | IC | ICIR | RIC | RICIR | AR | IR |
|---|---|---|---|---|---|---|
| LSTM | 0.050 | 1.837 | 0.053 | 1.814 | 0.830 | 2.879 |
| GRU | 0.042 | 1.832 | 0.047 | 1.817 | 1.159 | 4.387 |
| Transformer | 0.046 | 1.816 | 0.042 | 1.799 | 0.764 | 2.278 |
| GAT | 0.045 | 1.826 | 0.041 | 1.825 | 0.817 | 2.338 |
| GCN | 0.003 | 1.814 | 0.009 | 1.810 | 0.463 | 1.375 |
| DTML | 0.046 | 1.839 | 0.053 | 1.818 | 0.504 | 1.357 |
| XGBoost | 0.005 | 1.873 | 0.000 | 1.809 | 0.377 | 0.680 |
| Random Forest | 0.002 | 1.103 | 0.003 | 1.865 | 0.181 | -0.071 |
| MASTER | 0.059 | 0.390 | 0.067 | 0.430 | 0.290 | 2.400 |
| FactorVAE | 0.030 | 1.844 | 0.026 | 1.831 | 0.519 | 1.598 |
| StockMixer | 0.031 | 0.220 | 0.029 | 0.204 | 0.093 | 1.939 |
| GrifFinNet | 0.081 | 1.844 | 0.081 | 1.822 | 1.278 | 3.937 |

GrifFinNet achieves strong and consistent results across both benchmarks. For ranking metrics, the model attains an IC of 0.081 on CSI300 and 0.097 on CSI800, confirming robust predictive ability for cross-sectional returns. The ICIR scores reach 1.844 on CSI300 and 1.892 on CSI800, while RICIR values achieve 1.822 and 1.856 respectively, demonstrating stable prediction reliability across different market conditions.

In portfolio-level evaluation, the model achieves positive performance across both benchmarks with an annualized excess return (AR) of 1.278 and information ratio (IR) of 3.937 on CSI300, and 0.913 and 1.198 respectively on CSI800. While absolute performance varies between indices, both results demonstrate the model's consistent effectiveness in practical investment contexts.

Among the recent benchmark models, MASTER, StockMixer, and FactorVAE provide representative comparisons. MASTER employs market-guided gating to integrate intra- and inter-stock dependencies but underperforms on our datasets. StockMixer, while lightweight and efficient, struggles to capture complex market dynamics. FactorVAE achieves reasonable risk estimation but shows weaker predictive accuracy (IC = 0.030).

By contrast, GrifFinNet outperforms all baselines, achieving an IC of 0.081 (a 37.3% improvement over MASTER (0.059)) and strong ICIR and RICIR values. In profitability, GrifFinNet attains an AR of 1.278, exceeding the next best model (GRU, 1.159) by ~10%. Its IR of 3.937 further underscores its superior risk-adjusted performance.

Overall, these results confirm that the integration of multi-relational graph modeling with temporal encoding enables GrifFinNet to capture inter-stock dependencies more effectively than both sequential and graph-only baselines.

## Ablation Study

To assess the contribution of each core component, we conduct systematic ablation experiments on both CSI300 and CSI800. Table 1 examines model variants by removing individual modules in turn, i.e., gating, multi-relational modeling, attention, or temporal encoding.

Table 4. Ablation results on CSI300 and CSI800.

|  | Configuration | IC | ICIR | RIC | RICIR | AR | IR |
|---|---|---|---|---|---|---|---|
| CSI300 | GrifFinNet | 0.081 | 1.844 | 0.081 | 1.822 | 1.278 | 3.937 |
|  | w/o gating | 0.053 | 1.851 | 0.048 | 1.831 | 0.885 | 1.640 |
|  | w/o relation | 0.055 | 1.857 | 0.049 | 1.831 | 0.880 | 1.593 |
|  | w/o attention | 0.055 | 1.830 | 0.052 | 1.829 | 1.028 | 2.265 |
|  | w/o encoder | 0.030 | 1.867 | 0.026 | 1.849 | 0.489 | -0.161 |
| CSI800 | GrifFinNet | 0.097 | 1.837 | 0.097 | 1.817 | 0.913 | 1.198 |
|  | w/o gating | 0.061 | 1.864 | 0.083 | 1.852 | 1.065 | 2.114 |
|  | w/o relation | 0.058 | 1.858 | 0.080 | 1.836 | 1.099 | 2.302 |
|  | w/o attention | 0.063 | 1.835 | 0.079 | 1.827 | 1.216 | 3.212 |
|  | w/o encoder | 0.063 | 1.832 | 0.068 | 1.799 | 1.011 | 1.677 |

Removing the gating module causes an IC drop of 34.6% on CSI300 and 37.1% on CSI800, showing that dynamic regulation of relational information is essential for adapting to varying market conditions.

Eliminating industry and institutional graphs leads to significant decreases in IC (32.1% on CSI300 and 40.2% on CSI800), demonstrating that capturing heterogeneous inter-stock relationships is critical for accurate cross-sectional return prediction.

While some ablated configurations show improvements in portfolio-level metrics, these gains come at the expense of reduced predictive precision.

Attention contributes meaningfully to predictive accuracy, with IC reductions of 32.1% (CSI300) and 35.1% (CSI800) when omitted. Effects on portfolio returns vary: while CSI800 shows slight gains in AR and IR without attention, CSI300 exhibits sharp decreases, underscoring the mechanism's role in stabilizing performance across markets.

The strongest degradation occurs when temporal encoding is removed: IC falls by 63.0% on CSI300 and 35.1% on CSI800, and IR becomes negative on CSI300. This highlights the indispensability of temporal modeling in financial time series, where sequential dependencies are central to return dynamics.

The ablation study demonstrates that GrifFinNet's performance depends on the synergistic integration of all major components. GrifFinNet consistently achieves the highest IC and RIC scores across both datasets, confirming superior cross-sectional return prediction accuracy. While some ablated configurations show marginal improvements in information ratios or portfolio returns, these gains come at the cost of reduced predictive accuracy. The complete model strikes an optimal balance between prediction precision and portfolio performance, prioritizing the fundamental objective of accurate return forecasting over potentially unstable portfolio-level optimizations.

## Parameter Sensitivity Analysis

Table 5. Sensitivity of GrifFinNet to key hyperparameters on CSI300 and CSI800. Configurations include model dimension (dim), number of attention heads (heads), dropout rate (dropout), and number of Transformer layers (layers). The main model GrifFinNet denotes the optimal configuration.

|  | Configuration | IC | ICIR | RIC | RICIR | AR | IR |
|---|---|---|---|---|---|---|---|
| CSI300 | GrifFinNet | 0.081 | 1.844 | 0.081 | 1.822 | 1.278 | 3.937 |
|  | dim_64 | 0.058 | 1.829 | 0.065 | 1.840 | 0.811 | 1.659 |
|  | dim_128 | 0.033 | 1.835 | 0.029 | 1.832 | 0.434 | -0.434 |
|  | dim_512 | 0.023 | 1.802 | 0.015 | 1.786 | 0.537 | 0.087 |
|  | heads_1 | 0.059 | 1.834 | 0.057 | 1.825 | 0.935 | 1.900 |
|  | heads_2 | 0.058 | 1.834 | 0.054 | 1.820 | 0.879 | 1.592 |
|  | heads_4 | 0.071 | 1.861 | 0.065 | 1.831 | 0.989 | 2.064 |
|  | dropout_0.2 | 0.051 | 1.838 | 0.047 | 1.813 | 0.928 | 1.904 |
|  | dropout_0.25 | 0.056 | 1.847 | 0.050 | 1.834 | 0.891 | 1.631 |
|  | dropout_0.3 | 0.057 | 1.840 | 0.048 | 1.844 | 0.850 | 1.418 |
| | layers_2 | 0.060 | 1.827 | 0.056 | 1.817 | 1.100 | 2.946 |
| | layers_4 | 0.087 | 1.817 | 0.083 | 1.821 | 0.875 | 1.644 |
| | layers_6 | 0.088 | 1.834 | 0.083 | 1.814 | 0.803 | 1.416 |
| | GrifFinNet | 0.097 | 1.837 | 0.097 | 1.817 | 0.913 | 1.198 |
| | dim_64 | 0.074 | 1.841 | 0.077 | 1.821 | 0.919 | 1.426 |
| | dim_128 | 0.072 | 1.841 | 0.074 | 1.828 | 0.798 | 0.541 |
| | dim_256 | 0.058 | 1.858 | 0.080 | 1.836 | 1.099 | 2.302 |
| | heads_1 | 0.053 | 1.845 | 0.072 | 1.824 | 1.154 | 2.667 |
| | heads_2 | 0.055 | 1.851 | 0.074 | 1.835 | 1.278 | 3.557 |
| CSI800 | heads_4 | 0.055 | 1.853 | 0.073 | 1.843 | 1.123 | 2.529 |
| | dropout_0.2 | 0.055 | 1.872 | 0.076 | 1.834 | 1.156 | 2.768 |
| | dropout_0.25 | 0.050 | 1.866 | 0.070 | 1.825 | 1.095 | 2.310 |
| | dropout_0.3 | 0.057 | 1.857 | 0.072 | 1.827 | 1.175 | 2.751 |
| | layers_2 | 0.062 | 1.849 | 0.084 | 1.827 | 1.136 | 2.635 |
| | layers_4 | 0.061 | 1.856 | 0.078 | 1.847 | 1.212 | 2.838 |
| | layers_6 | 0.073 | 1.849 | 0.088 | 1.825 | 1.207 | 2.997 |

To assess the robustness of GrifFinNet, we conduct sensitivity experiments on several key hyperparameters (Table 5). All experiments vary one parameter at a time while keeping others fixed at their optimal settings. Results are reported separately for CSI300 and CSI800.

We first examine the effect of the embedding dimension (dim). For CSI300, the best performance occurs at dim=256. Lower dimensions (64, 128) substantially degrade IC, while a higher dimension (512) also reduces accuracy, indicating over-parameterization. For CSI800, which covers a larger and more complex stock universe, dim=512 yields the highest IC, outperforming all lower settings. This confirms that dataset scale should guide the choice of model capacity.

We next vary the number of attention heads. On CSI300, four heads achieve results closest to the main model (3 heads) but still lag in IC, while one or two heads cause more severe drops. On CSI800, all the other tested variants (1, 2, or 4 heads) underperform relative to the main configuration (3 heads). These findings suggest that the chosen attention head configuration in the main model most effectively captures multi-scale dependencies.

Dropout values of 0.2, 0.25, and 0.3 all result in lower IC and declines in AR and IR compared to the value dropout=0.15 in GrifFinNet. This pattern holds across both CSI300 and CSI800, indicating that the adopted dropout level 0.15 strikes the best balance between preventing overfitting and preserving predictive strength.

Increasing the number of Transformer layers provides only marginal IC improvements but at the cost of significant declines in portfolio metrics (AR and IR). For example, in CSI300, four and six layers slightly raise IC but reduce AR by 31.5% and 37.1%, respectively. Similar trade-offs occur in CSI800. These results highlight that a shallow architecture achieves a better balance between accuracy and investment performance.

The sensitivity analysis demonstrates that GrifFinNet is robust to moderate hyperparameter variation, but performance is sharply affected by model dimension and temporal depth. The chosen configurations, i.e., dim=256 for CSI300 and dim=512 for CSI800, one Transformer layer, eight attention heads, and a dropout rate of 0.15, represent well-calibrated settings that balance predictive accuracy with portfolio stability.

## Contribution of Multi-Relational Modeling

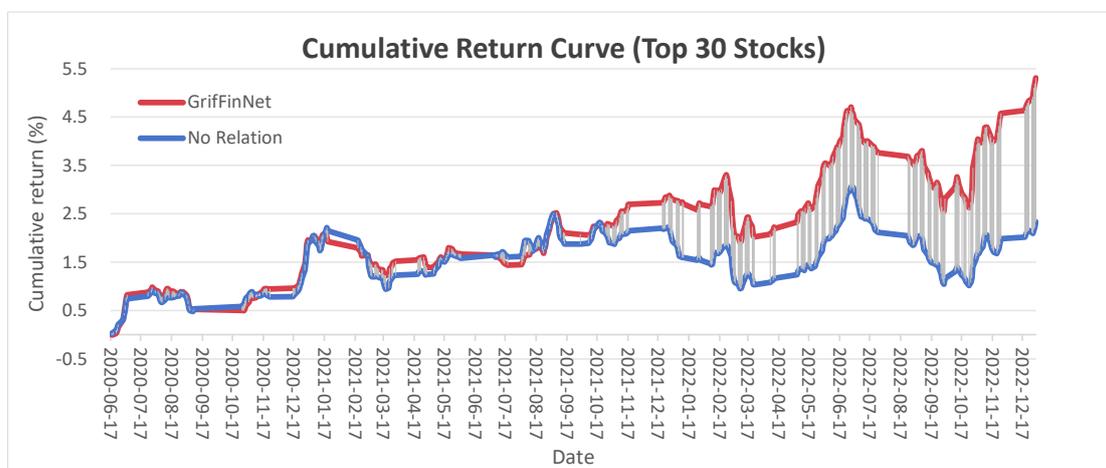

(a)

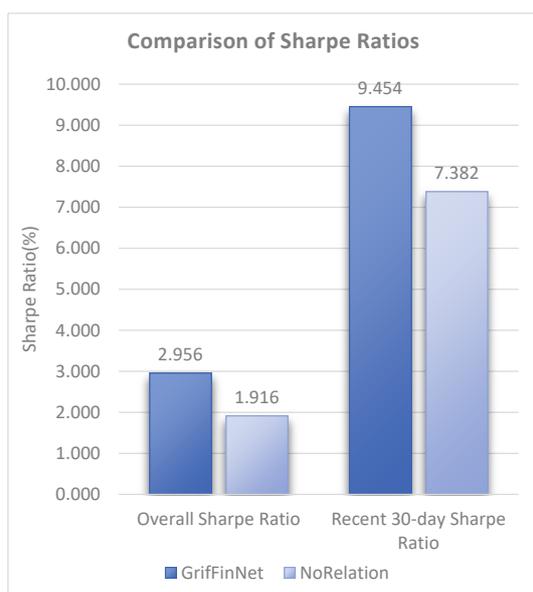
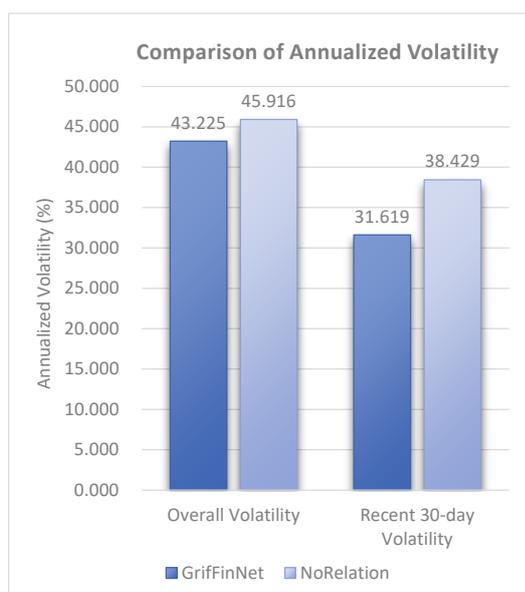

(b) (c)

Figure 2. Evaluation of the multi-relational modeling module. (a) Cumulative return comparison between GrifFinNet and its ablation model without multi-relational modeling on the CSI300 dataset. (b) Comparison of overall and 30-day Sharpe ratios for GrifFinNet and the ablation model on the CSI300 dataset. (c) Annualized volatility of GrifFinNet and the ablation model on the CSI300 dataset.

GrifFinNet demonstrates clear advantages over its ablation variant without multi-relational modeling. As shown in Figure 2 (a), cumulative return curves highlight the practical value of incorporating industry and institutional relations. By modeling both types of dependencies, GrifFinNet captures richer market structures and achieves a final cumulative return of 386.8%, compared to only 162.9% for the ablation model. This 137% improvement underscores the central role of multi-relational information in stock return prediction.

To assess risk-return trade-offs, we evaluate Sharpe ratios across both the entire testing period and recent 30-day windows. As shown in Figure 2 (b), GrifFinNet consistently achieves higher Sharpe ratios: 2.956 overall versus 1.916 for the ablation model, and 9.454 versus 7.382 for short-term performance. These results confirm that multi-relational modeling improves not only profitability but also the stability of returns across different horizons.

We further compare annualized volatility to assess risk control. Figure 2 (c) shows that GrifFinNet achieves lower volatility (43.225%) than the ablation model (45.916%), a reduction of 5.9%. This indicates that multi-relational modeling enhances both return generation and risk management in portfolio-level applications.

## Contribution of the Adaptive Gating Mechanism

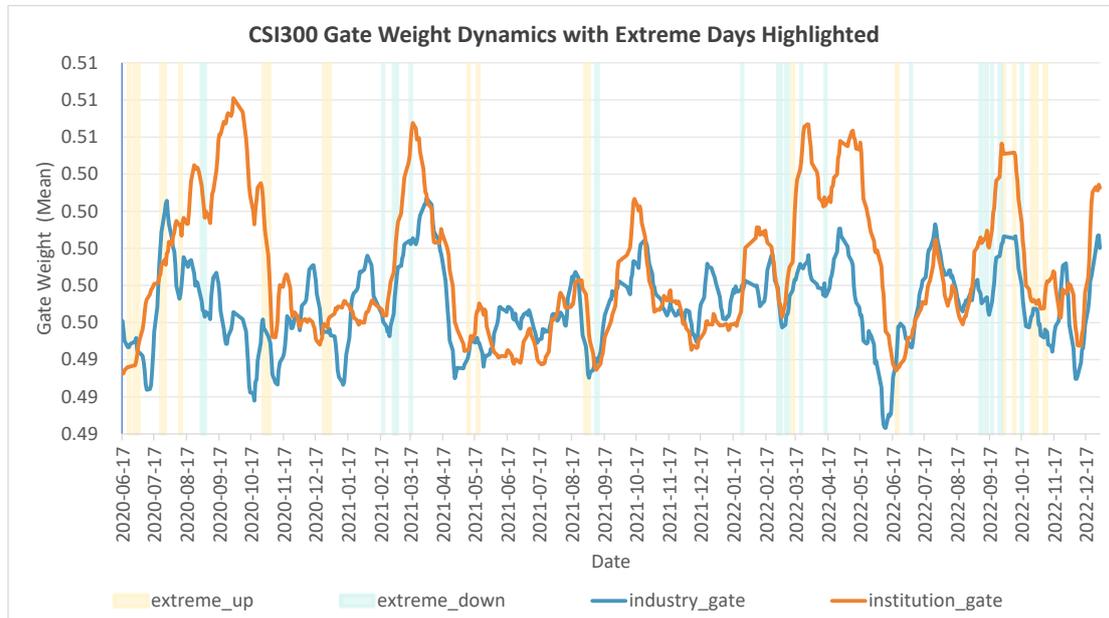

(a)

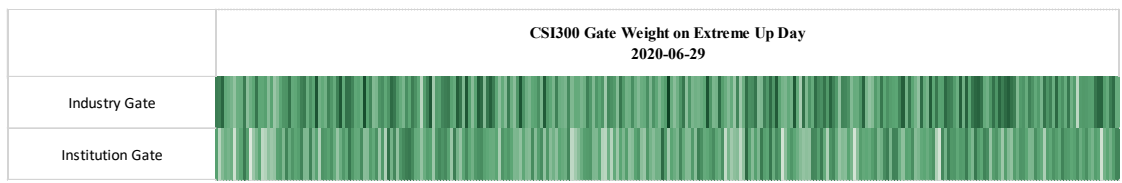

(b)

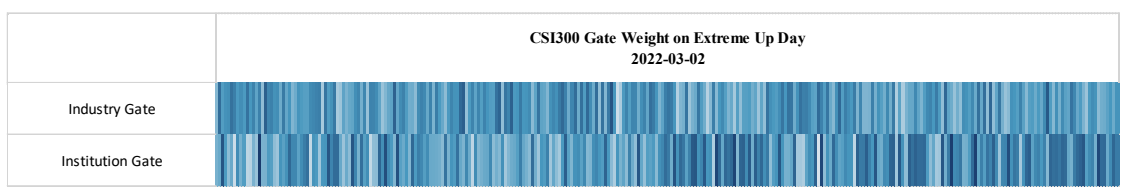

(c)

Figure 3. Evaluation of the adaptive gating mechanism. (a) Dynamics of gating weights during extreme market days. Gating weight distributions on extreme up (b) and extreme down (c) market days.

Beyond multi-relational modeling, GrifFinNet employs an adaptive gating mechanism to dynamically regulate the importance of industry and institutional relationships (Figure 3 (a)). This mechanism adjusts feature weights in real time according to market conditions, and enables the model to prioritize the most relevant information sources. On extreme market days (identified as the top and bottom 5% of returns), the gating mechanism exhibits clear patterns: it increases reliance on industry relations during market upswings, while emphasizing institutional relations in downturns, reflecting risk-averse behavior.

To further illustrate this adaptability, Figure 3 (b) and (c) present heatmaps of gating weight distributions on a representative extreme up day (June 29, 2020) and an extreme down day (March 2, 2022). The contrasting patterns confirm that the gating mechanism effectively reallocates attention under different regimes, and contribute to the robustness of GrifFinNet across varying market conditions.

Together, these findings highlight the complementary roles of multi-relational modeling and adaptive gating. While multi-relational graphs provide a richer structural foundation, the gating mechanism enables real-time adjustment to evolving market conditions. Their integration allows GrifFinNet to deliver superior predictive accuracy, higher profitability, and improved risk-adjusted performance.

## Conclusions

This study introduced GrifFinNet, a graph-relation integrated Transformer framework for stock return prediction. By combining heterogeneous inter-stock graphs based on industry affiliation and institutional ownership with a temporal Transformer encoder, GrifFinNet captures both structural dependencies and sequential dynamics within financial markets. An adaptive gating mechanism further enhances the framework by dynamically weighting heterogeneous relations in response to shifting market conditions.

Experiments on two major Chinese A-share indices (CSI300 and CSI800) demonstrate that GrifFinNet consistently outperforms a broad range of sequential, graph-based, and machine learning baselines across both predictive accuracy and portfolio performance. In addition, the adaptive fusion of relational information provides interpretable insights into market behavior, making the framework valuable for both prediction and analysis.

While GrifFinNet delivers strong empirical results, several limitations remain to be resolved in the future studies. 1) Only industry and institutional relations were modeled, and this study did not consider other potentially influential relationships such as investor sentiment, supply-chain dependencies, and geographical clustering. 2) The focus of this study was primarily short-term prediction. This restricts the applicability of the framework for long-horizon investment strategies where different dynamics may dominate. 3) Although the adaptive gating mechanism introduces a degree of interpretability by revealing shifts in relational importance, the model still lacks fine-grained interpretability at the level of individual investment decisions.

Future work will address these limitations in several directions. Expanding the relational space to incorporate sentiment, macroeconomic linkages, and global market influences may yield richer relational graphs. Extending the framework to multi-horizon prediction tasks could enhance its adaptability for both short-term trading and long-term asset allocation. Finally, integrating advanced interpretability techniques (such as counterfactual reasoning or attribution-based interpretability) would allow deeper understanding of the mechanisms driving model outputs, making the approach more actionable for decision-makers.

## Acknowledgements

This work was supported by the National Natural Science Foundation of China (No. 82574200), Development Project of Jilin Province of China (No. 20220508125RC), Guizhou Provincial Science and Technology Projects (ZK2023-297), the Science and Technology Foundation of Health Commission of Guizhou Province (gzwkj2023-565), and the Fundamental Research Funds for the Central Universities (JLU).

## Conflicts of Interests

The authors declared no conflicts of interest.